\documentclass{iau}
\usepackage{graphicx}

\newcommand{\dd}{\mathrm{d}}

\title[ Variation of bulk Lorentz factor in AGNs jets] %% give here short title %%
{Influence of an AGN complex photon field on the jet bulk Lorentz factor through Compton rocket effect}

\author[Thomas Vuillaume, Gilles Henri \& Pierre-Olivier Petrucci]   %% give here short author list %%
{Thomas Vuillaume$^1$,
Gilles Henri$^1$ \and Pierre-Olivier Petrucci$^1$}

\affiliation{Universit\'e Grenoble Alpes, IPAG, F-38000 Grenoble, France \& 
	CNRS, IPAG, F-38000 Grenoble, France,
\\ email: {\tt thomas.vuillaume@obs.ujf-grenoble.fr} \\[\affilskip]}

\pubyear{2014}
\volume{313}  %% insert here IAU Symposium No.
\pagerange{}
\jname{Extragalactic jets from every angle}
\editors{}
\begin{document}

\maketitle

%\begin{abstract} 
%Despite many studies, the jets acceleration to relativistic speeds is still misunderstood. In this work, we study the acceleration of hot plasma to relativistic speed through the Compton rocket effect which is viable in the two-flow paradigm. We compute the equilibrium bulk Lorentz factor along the jet in the complex photon field of an AGN which is composed of different external sources (accretion disk, dusty torus and broad line region). We show that this Lorentz factor will be subject to changes along the jet, corresponding to accelerating and decelerating zones of the flow and discuss some observational implications.

               
%\end{abstract}

%\cite[Anders \& Zinner (1993)]{AndersZinner93} 

\firstsection % if your document starts with a section,
\section{Introduction}

It is now widely admitted that AGN's jets hold relativistic flows. First evidences go back to the 70's with the observation of superluminal motions (\cite{1971ApJ...170..207C}) which are only possible for actual speeds of $0.7c$ at least.
 However, a lot of questions on the speed of these flows remain. Mainly, we still do not know the mechanism driving them to relativistic speeds or neither do we know the spatial distribution of these speeds in the flows. They can be characterized by their bulk Lorentz factor $\Gamma_b=\left( 1- \beta_b^2\right)^{-1/2}$ rather than their speed $V_b$ with $\beta_b=V_b/c$. In some studies, the distribution of Lorentz factors appears to follow power laws with an accelerating and/or a decelerating phase (\cite[Marscher 1980]{Marscher:1980js},\cite{Ghisellini:1985wr}, \cite{Boutelier01102008}).

Our work takes place in the two-flow paradigm (\cite{1989MNRAS.237..411S}) where the jet is composed of a mildly relativistic sheath, filled with $e^- / p^+$ and an ultra-relativistic spine composed of $e^- / e^+$ pairs responsible for most of the emission. The outer jet acts as an energy reservoir for the particles of the spine, which will be continuously thermalized along the jet via the second order Fermi process. This is in agreement with diffuse X-ray emission observed in FRI which favors a distributed particle acceleration rather than localized shocks (\cite{2007ApJ...670L..81H}). In this paradigm, the plasma is subject to the Compton rocket effect which will naturally drive the flow to relativistic speeds.

\section{$\Gamma_{b}$ \& equilibrium}

\begin{figure}[h]
\begin{center}
\includegraphics[width=2.4in,trim= 5cm 2.5cm 5cm 0cm,clip=true]{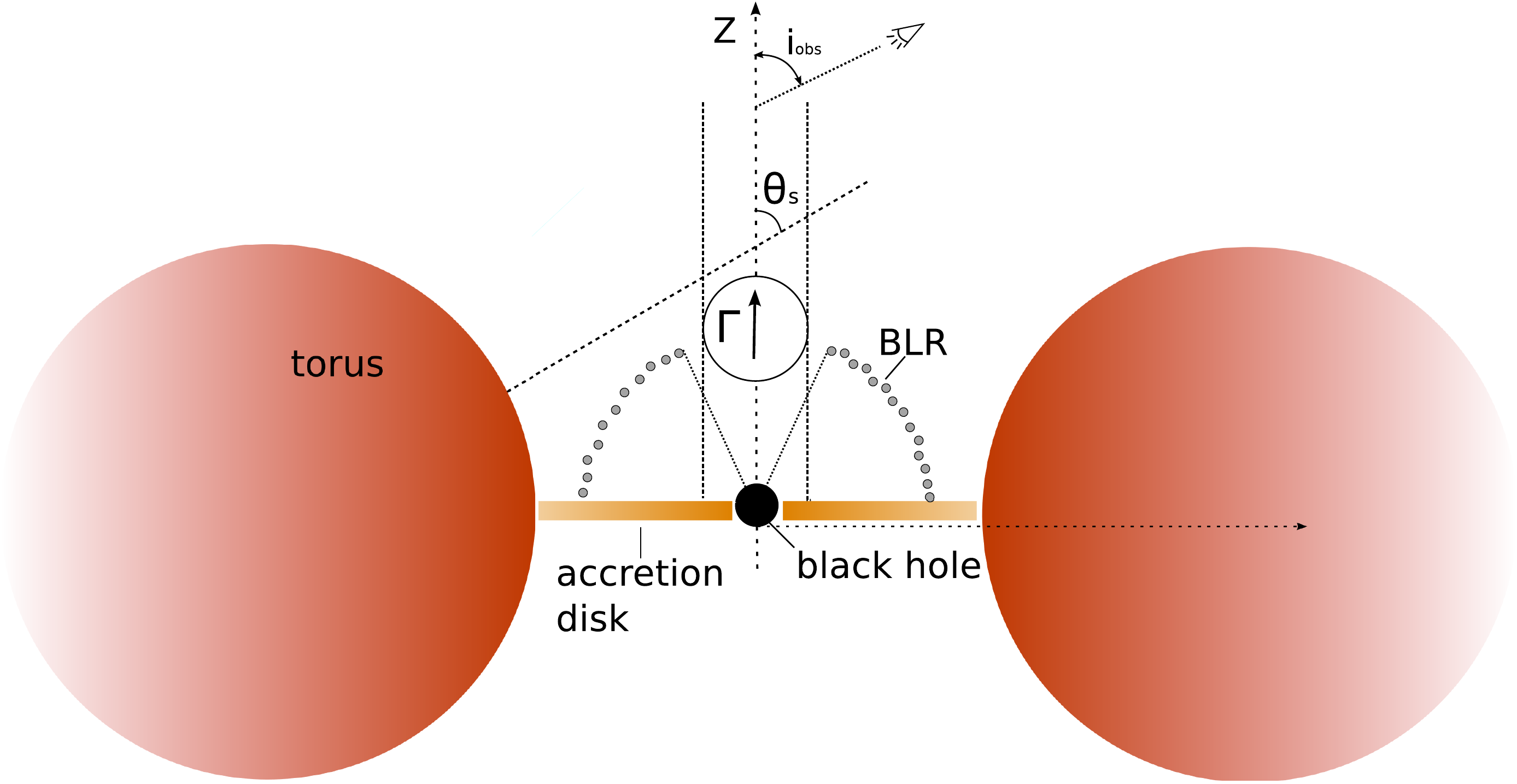}
\caption{\label{fig:sketch}The big picture: sketch edge on of the global model geometry (not to scale) with the accretion disk, the dusty torus and the BLR.}
\end{center}
\end{figure}

\cite[O'dell, 1981]{1981ApJ...243L.147O} showed that "a plasma of relativistic particles exposed to an anisotropic radiation field acts as a rocket - a Compton rocket" because of the reaction force imposed by the inverse Compton radiation from the particles. In the Thomson regime, this force is proportional to the flux in the plasma rest frame $ \displaystyle H^* = \frac{1}{4\pi}\int I^*_{\nu_s} (\Omega^*_s, \Gamma_b)  \cos\theta^*_s \dd \Omega^*_s \dd \nu^*_s $ with the quantities in the rest frame: $I^*_{\nu_s}$ the specific intensity of the radiation reaching the plasma with a solid angle $\dd \Omega^*$ and an angle $\theta^*_s$ defined figure \ref{fig:sketch} at a frequency $\nu^*_s$. 
Thus, the bulk of particles reaches an equilibrium velocity, which can be represented by the equilibrium bulk Lorentz factor $\Gamma_{eq}$, when $H^* = 0$.
To compute the equilibrium bulk Lorentz factor, we need to compute precisely the external photon field at any position of the plasma. Here, we model the AGN including three main sources of soft photons: a standard accretion disk, a dusty torus in thermal equilibrium and a broad line region (BLR) modeled as a spherical shell of clouds (see figure \ref{fig:sketch})

\section{Evolution of $\Gamma_{eq}$ along the jet}

\begin{figure}[h]
% \vspace*{-2.0 cm}
\begin{center}
 \includegraphics[width=3.4in]{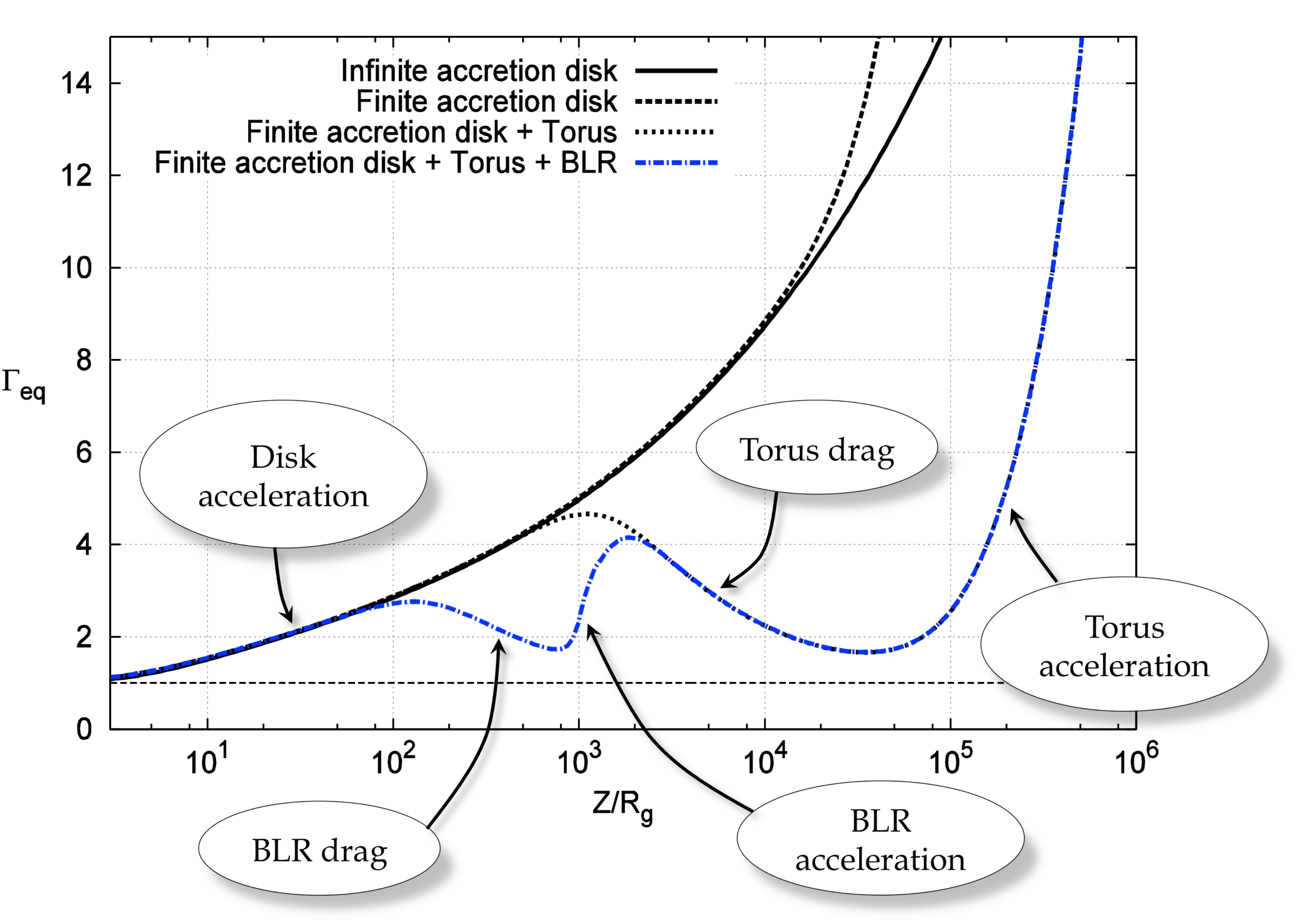} 
% \vspace*{-1.0 cm}
 \caption{\label{fig:gam_eq} $\Gamma_{eq}$ resulting of the Compton rocket effect for different external photon sources. The geometry is described in figure \ref{fig:sketch} with the following parameters: finite and infinite accretion disk have an inner radius $R_{in}=3 R_g$. The finite disk has an outer radius $R_{out}=5 \times10^4 R_g$. $D_{torus} =10^5 R_g$, $R_{torus} = 5 \times10^4 R_g$, $R_{BLR} = 10^3 R_g$}
\end{center}
\end{figure}

Figure \ref{fig:gam_eq} represents $\Gamma_{eq}$ for different configurations of external sources. Photons from the accretion disk are emitted upward which corresponds to a positive flux in the bulk rest frame $H^* > 0$. This leads to an inverse Compton emission backward and thus a reaction force forward, which at the end is accelerating the flow ($\Gamma_{eq}$ increases). However, because of aberration effects, the situation is more complex for the dusty torus and the BLR. Until a certain altitude, photons from the dusty torus or from the BLR generate a negative rest frame flux, $H^* < 0$ which produces a backward force, or Compton drag, decelerating the flow. It is only when $H*>0$ that the flow is accelerated again.

\end{document}